# Exploring the Relationship Between "Positive Risk Balance" and "Absence of Unreasonable Risk"

Francesca Favarò

**Introduction**

International discussions on the overarching topic of how to define and quantify what a "safe enough" Automated Driving System (ADS) is are currently hinged on the question of determining the relationship between "positive risk balance" (PRB) and "absence of unreasonable risk" (AUR).

ISO 5083 - *Safety for automated driving systems*, currently under development from ISO TC 22 SC 32 WG 13, is expected to provide guidelines for the structure that a safety case for ADS could follow, stemming from considerations of both AUR and PRB [1]. At this stage, discussions in the international fora have not yet reached consensus on what the relationship is between the two concepts, and whether PRB should indeed be considered in and of itself a safety goal (similar to AUR), or whether PRB is just an instantiation of a more general goal of a different nature. This paper will thus attempt to provide an interpretation of these concepts, with the hope to further the conversation on these important safety topics. To that end, this paper will start with an overview of the notions of PRB and AUR; it will then summarize different positions of the present debate; finally, it will conclude that two possible interpretations exist for PRB, and that failure to distinguish them can lead to misunderstanding different parties' positions. The argumentation in this paper is aimed at showing that the two interpretations for PRB can actually complement each other, but can be considered independently, and can both be subsumed within *non-prescriptive* guidelines toward ADS safety assurance.

**Positive Risk Balance (PRB) Notion Development**

The first appearance of the notion of PRB for ADS is often attributed to a June 2017 report from the German Ethics Commission, which states (in both its full-length version [2], and abridged report [3]):

> "The protection of individuals takes precedence over all other utilitarian considerations. The objective is to reduce the level of harm until it is completely prevented. The licensing of automated systems is not justifiable unless it promises to produce at least a diminution in harm compared with human driving, in other words **a positive balance of risks**" *[2,3]* (emphasis added)

This notion was later reworked by the European Union report on "Ethics for Connected and Automated Vehicles" [4], which both adapted the general notion of PRB and added emphasis on the ethical principles one should consider when debating certification of those vehicles. The EU report makes the following summary recommendation:

*"In line with the principle of non-maleficence, a minimal requirement for manufacturers and deployers is to ensure that CAVs (connected automated vehicles) **decrease, or at least do not increase, the amount of physical harm** incurred by users of CAVs or other road users that are in interaction with CAVs, **compared to the harm that is inflicted on these groups by an appropriately calculated benchmark based on conventional driving**. A further requirement, in line with the principle of justice, is that **no category of road user [...] should end up being more at risk of harm from CAVs than they would be against this same benchmark**" [4]* (emphasis added)

The recommendation quoted above is distilled from a number of principles that can be found in the EU report, including the "non-maleficence" principle of *primum non nocere* (translation: first do no harm[1]), and, secondly, of "beneficence", that is, of positively contributing to the welfare of all individuals. While neither the recommendation nor the EU principles mention PRB directly, the EU report does reference the German Ethics Commission work, and the verbiage reported above has been tied directly to PRB within various international debates.

The first formal definition of PRB came in 2019, with the "[Safety First for Automated Driving](#)" (SaFAD) white paper [5], which states the following:

*"Positive Risk Balance is the result of a risk benefit evaluation with a lower remaining risk of traffic participation due to automated vehicles. This includes the fact that automated vehicles **causes** [sic] **less crashes on average compared to the average human driver**" [5]* (emphasis added)

Such definition was further reworked in 2020 in the informative ISO Technical Report 4804 [6], which is considered as the starting point for the new Technical Specification ISO 5083, expected for 2023. The ISO TR 4804:2020 definition reads as follows:

*"PRB: benefit of sufficiently mitigating residual risk of traffic participation due to automated vehicles. Note 1: This includes the **expectation that automated vehicles cause less crashes on average compared to those made by drivers**. Note 2: PRB is one of the concepts that can be considered when defining the acceptance criteria of ISO PAS 21448:2019 ([7])" [6]* (emphasis added)

**The Notion of Absence of Unreasonable Risk (AUR)**

The notion of "absence of unreasonable risk" is grounded in over a decade of safety standardization activities. While its original formulation is derived from a number of previous standards, the experts community refers to ISO 26262:2018 definitions [8], which are considered mature, stable, and state of the art. For example, there is evidence of them being followed and harmonized in both regulatory drafts seen from the EU and individual States in Europe as well as in various industry reports - see for instance [5, 9, 10]. The following definitions are required to ensure an understanding of the debate happening internationally, and ground the definition of AUR highlighted below:

- **Risk:** combination of the probability of occurrence of harm and the severity of that harm [8]
- **Unreasonable risk**: risk judged to be unacceptable in a certain context according to valid societal moral concepts [8]
- **Acceptable**[2]: Sufficient to achieve the overall item risk as determined in the safety case [12]
- **Acceptable risk**: overall risk sufficient to satisfy the safety case [11] (adapted from [12] definition of "acceptable")
- **Safety**: **absence of unreasonable risk** [8]
- **Safety case (1)**: argument that (functional)[3] safety is achieved for items or elements and satisfied by evidence compiled from work products of activities during development [8]
- **Safety case (2)**: a structured argument, supported by a body of evidence, that provides a compelling, comprehensible and valid case that a system is safe for a given application in a given environment [12]

---

[1] From the PRB perspective, this is not to be necessarily intended as an absolute tenet against physical harm, but rather as the need to not worsen the status-quo, which would harm public welfare.

**The Debate**

The current international debate can be summarized in two main questions:

- What is the relationship between AUR and PRB?
- How would one quantify "positive risk balance"?

Bringing this one step further, given that safety itself has been defined as the absence of unreasonable risk (i.e., based on definitions agreed upon and with established consensus from [8], one has that *Safety = AUR* for different safety domains), can PRB be used instead to quantify the *"enough"* portion of the question?

Under that perspective, it is clear that the two questions are highly intertwined with each other and go at the heart of determining what role PRB would have in the assessment of the "safe enough" question.

**The Contributions and Their Interpretations**

At present, international experts have provided input and expressed opinions that could be summarized in the following way:

---

[2] No definition for "unacceptable" was found.
[3] ISO 26262 includes a note in this definition, which states that "functional safety" within the definition can be replaced with other kinds of safety as well. This is why the word "functional" was placed here in parentheses.

- The notion of PRB is grounded in the two ethical principles of nonmaleficence and beneficence recognized by the European Union. As such, it can inform the societal moral concepts that ground the notion of "unreasonable risk".
- PRB attempts at making more explicit the balance between two different aggregate levels of risk, one achieved with a transportation system that includes ADS and one that does not include them. While a comparison may be appropriate, safety practices grounded in the notion of AUR would require that both sides of any drawn risk balance be deemed "reasonable".
- In terms of its quantification, PRB can be tied to a benchmarking requirement set on the frequency of incurred harm (e.g., fatality or other agreed-upon severity level). This approach, in its many variations, can be formulated with the simple notation:

**Frequency of the occurrence of (certain level of) harm < = Threshold value     (1)**

It is the author's opinion that it is actually the use of such an intuitive equation that can lead to possible misunderstandings and lack of consensus in the international debate. It is in fact argued here that two possible approaches and interpretations of PRB are possible, and that they inform different parts of the equation above, as well as apply it at different stages within the ADS development process.

**Interpretation #1:** The first interpretation focuses on the evaluation of the balance across the two terms of equation (1) through the use of PRB as an *outcome-based approach* that sets a requirement on the ADS aggregate performance. This interpretation, more "literal" in its reading of the German Ethics report [2] and the EU ethics guidelines [4], would inform safety assurance processes for the *completed* system. Similar to how it was presented in SaFAD [5], this would place PRB as a high-level *goal to ensure the realization of the full safety benefits* obtainable through the deployment of an ADS.

**Interpretation #2:** The second interpretation focuses instead on the "acceptability" of the threshold value, and employs PRB as a means to guide the ADS development process starting from its conceptual phases and from the vehicle-level hazard analysis and risk assessment. In other words, this second approach interprets PRB as a *risk criterion used for the evaluation of acceptable risks tied to vehicle-level hazards*, as informed by more traditional hazard analysis and functional safety. For example, Warg and co-authors [13] bring forth the idea of a quantitative risk norm, viewed as a "budget" for defining a threshold for acceptable risk (defined through the combination of frequency and severity of outcomes) during *development* stages. Within this perspective, PRB is referred to as a method to determine what level of risk is actually tolerable, which is a different perspective than the more holistic performance assessment originally presented within the German ethics reports [2]. Under this perspective, PRB would become the chosen criterion (out of many other possible ones) for defining unacceptable risk, which makes it (by the definition above from [8]) the criterion for defining unreasonable risk.

While there are clear synergies in the two interpretations described above, one could argue that failure to recognize their differences could lead to a dichotomy of understanding between different individuals (and perhaps countries) that have traditional and historical safety ties more aligned with one or another approach. Conversely, the recognition of such differences makes it possible to understand that the two PRB interpretations can complement each other, and that they could both fit within non-prescriptive guidelines and ADS safety considerations for deployment. While the first one would be considered a *safety assurance metric for a completed system*, the second one could be used to *guide development of a safe system from the starting conceptual phases*. Furthermore, the *two uses could also be employed independently*, so that *leveraging PRB within the first approach for safety assurance would not necessarily constrain a user to employ PRB as their stated risk criterion for acceptability*.

The safety literature abounds with other methodologies that can also inform the "threshold value" selected. Some, like PRB, remain open to a possible use according to both interpretations; others, instead, only make sense within functional safety processes. Example methods, encountered in both regulatory drafts and industry proposals (see for example [10], [12], and [14]) and stemming from past safety literature exported from various domains of application, include:

- GAMAB = globalement au moins aussi bon (translation: as a whole at least as good as), which would place the threshold (though open to interpretation) to the current state of the art on the road (once more applicable across different crash severities). This principle has often been employed in the French railway industry to ensure that new proposals for guided transportation would be at least as good as *previous versions* (see Annex D of [15]). Similar to PRB (and easily alignable with it), GAMAB is also open to both interpretations.
- ALARP/ALARA = as low as reasonably practicable/achievable [15, 16], a criterion aligned with the second interpretation proposed above, which guides the implementation of hazard mitigations according to a consideration of practical and reasonable effort required to reduce tolerable risks into negligible regions[4].
- MEM = minimum endogenous mortality, which would set the threshold as a rate of fatalities per operational metric (e.g,. hours of operations, or miles of use). MEM has been used in a number of domains other than ADS safety for "first-ever" designs, meaning for novel products that did not have a readily available benchmark for risk comparison (most notably done for the first nuclear power plants in the US, and space launch vehicles in the aviation industry - see for instance, [17, 19, 20]).

To move the international standardization debate forwards within ISO, it is thus recommended that:

- The differences between the two PRB interpretations be clarified and consensus be reached on the definition of these approaches. Different wording and terminology could also be employed for the two interpretations, distinguishing the use of PRB as a goal within the first interpretation and as a method/risk criterion within the second one;

---

[4] Many different interpretations of the ALARP principle exist, with the core common denominator being that ALARP considerations come into effect after the risk level is below a given "intolerable" threshold. Some call this the "reducible region" of risk towards "tolerability", others call it a "tolerable region" towards "negligibility" [16, 17, 18].

- Methodologies needed to leverage PRB according to both the first and the second interpretation be further discussed.
    - In fact, it is the author opinion that PRB should be informatively considered as one of many possible approaches for the definition of benchmarks of interest (both holistically from an overall performance perspective as well as at the individual risk assessment level), where a number of factors (that would need to be specified by any given user within their safety case) can inform the type of benchmark considered. An example is provided in Table 1 below, where the selection of appropriate options could lead to selecting a benchmark that aligns with PRB or with other possible methodologies of interest.

Table 1. Identified "options" for benchmarks discriminants. Example only, not meant to be exhaustive

| Reference Area | Options |
| --- | --- |
| Benchmark Defining Factors - Data Sources | - Reference domain of data source: ODD-specific vs. aggregate (e.g., national) vs. ad-hoc use case<br>- Vehicle type: for example type/size/use similarity<br>- Technology Advancement: for example active safety features equipped in vehicles used for the benchmark<br>- Consequence: severity levels (with associated probability thresholds- e.g., MAIS), and/or possible inclusion of near-crashes (with defined metrics)<br>- Crash Type: breakdown by mechanisms vs. aggregate<br>- Other road user type: aggregate vs. categories of road users (e.g., pedestrians, vehicles, cyclists)<br>- Other road user status/skill: for example, attentive vs. distracted, or experienced/trained driver vs. novice |
| Evaluation Method | - Predictive (estimated) measure, with stated confidence interval<br>- Observed measure |
| Etiology (see [21]) | - Caused accidents only (initiator)<br>- Unavoided collisions in responders role |
| Performance Sough | - Level: neutral or positive<br>- Function: aggregate (overall) or competence-specific (scenario-based) |
| Evaluating Party | - Manufacturer<br>- Regulatory agency<br>- Third Party |

**References**

[1] Road vehicles — Safety for automated driving systems — Design, verification and validation. Approved Work Item ISO/AWI 5083. *Under development*

**About the author:**

Francesca Favarò is a Senior Researcher within the Safety Team at Waymo, an autonomous driving technology company, where she also coordinates external engagement with standards developing organizations. She is also an Associate Professor at San Jose State University's Department of Aviation and Technology and a research associate at the Mineta Transportation Institute. Dr. Favarò earned her Ph.D. and M.S. in Aerospace Engineering from Georgia Tech, and received her M.S. in Space Engineering and B.S. in Aerospace Engineering from Politecnico di Milano, Italy.